\documentclass[12pt]{article}

\usepackage[body={17.5cm, 23.1cm},right=2cm]{geometry}

\usepackage{mathtools}

\usepackage[cbgreek]{textgreek}

\usepackage{color}
\usepackage{graphicx}
\usepackage{epsf}
\usepackage{graphicx,epsfig}
\usepackage[mathscr]{euscript}
\usepackage{pbox}
\usepackage{float}
\usepackage{amsmath}
\usepackage{amssymb}
\usepackage{epsfig}
\usepackage{cite}
\usepackage{color,colordvi}
\newcommand{\be}{\begin{eqnarray}}
\newcommand{\ee}{\end{eqnarray}}
\newcommand{\bi}{\begin{itemize}}
\newcommand{\ei}{\end{itemize}}

\newcounter{hran}

\def\MSbar{\relax\ifmmode\overline{\rm MS}\else{$\overline{\rm MS}${ }}\fi}
\def\del{\partial}
\def\ta{\text{\textalpha}}

\begin{document}\thispagestyle{empty}

\vspace{0.5cm}

\def\thefootnote{\arabic{footnote}}
\setcounter{footnote}{0}

\def\s{\sigma}
\def\nn{\nonumber}
\def\p{\partial}
\def\ls{\left[}
\def\rs{\right]}
\def\lc{\left\{}
\def\rc{\right\}}
\def\S{\Sigma}
\def\l{\lambda}
\newcommand{\beq}{\begin{equation}}
\newcommand{\eeq}[1]{\label{#1}\end{equation}}
\newcommand{\bea}{\begin{eqnarray}}
\newcommand{\eea}[1]{\label{#1}\end{eqnarray}}

\renewcommand{\be}{\begin{eqnarray}}
\renewcommand{\ee}{\end{eqnarray}}
\renewcommand{\th}{\theta}
\newcommand{\bth}{\overline{\theta}}

\hspace*{12cm}
CERN-PH-TH/2014-134

\vspace*{1.2cm}

\begin{center}

{\Large \bf 
Higher Curvature Supergravity, 
 Supersymmetry Breaking and Inflation$^*$}
\\[1.5cm]
{\large   
Sergio Ferrara$^{a,b,c}$ and  Alex Kehagias$^{d,e}$ 
}
\\[0.9cm]

\vspace{.3cm}
{\normalsize {\it  $^{a}$  Physics Department, Theory Unit, CERN,
CH 1211, Geneva 23, Switzerland}}\\

 \vspace{.3cm}
{\normalsize {\it  $^{b}$  INFN - Laboratori Nazionali di Frascati 
\\
Via Enrico Fermi 40, I-00044 Frascati, Italy}}\\

 \vspace{.3cm}
{\normalsize {\it  $^{c}$  Department of Physics and Astronomy,
University of California \\ Los Angeles, CA 90095-1547, USA}}\\

\vspace{.3cm}
{\normalsize {\it  $^{d}$ Physics Division, National Technical University of Athens, 
15780 Zografou,
Athens, Greece}}\\

\vspace{.3cm}
{\normalsize { \it $^{e}$ Department of Theoretical Physics 
24 quai E. Ansermet, CH-1211 Geneva 4, Switzerland}}\\




\end{center}
\vspace{.8cm}
\hfill {\it \normalsize{dedicated to the memory of Bruno Zumino}}
\vspace{.9cm}

\begin{center}
{\small  \noindent \textbf{Abstract}} \\[0.5cm]
\end{center}
\noindent 
{\small
In these lectures, after a short introduction to cosmology,  we discuss the supergravity embedding of higher curvature models of inflation.
The
supergravity description of such models is presented for the two different formulations of minimal supergravity.  }
\vskip 2.5cm

\def\thefootnote{\arabic{footnote}}
\setcounter{footnote}{0}
\vfill
\vskip.2in
\line(1,0){250}\\
{\footnotesize {$^*$ 
Lectures given by S.Ferrara at the Erice ISSSP 2014, 52th Course, 24 June-3 July, 2014
}}




\baselineskip= 19pt
\newpage

\section{Introduction}

These lectures are devoted to the application of higher curvature supergravity to a particular class of cosmological models for inflation in which the ``inflaton" field is identified with the ``scalaron". The latter is a purely gravitational mode which arises when we add to the Einstein-Hilbert action a term quadratic in the (scalar) curvature 
\begin{eqnarray}
{\cal L}_{modified}={\cal L}_{EH}+\ta R^2=\frac{1}{2\kappa^2}R+\ta R^2, 
\end{eqnarray} 
where $\kappa^2=8\pi G=M_P^{-2}$. 
This theory is ``dual" to standard Einstein gravity coupled to a scalar field.
The revival of these models was motivated by the fact that the recent experiments Planck \cite{Planck} and  BICEP2 \cite{BICEP} seem to favor simple one-field cosmological models for inflation even if there is a tension between the two  experiments. In fact, while for the slow-roll parameter
\begin{eqnarray}
 n_S=1-6 \epsilon+2\eta\approx 1-\frac{2}{N}\approx 0.96 
 \end{eqnarray} 
i.e., the spectral index of scalar perturbations, the same formula agrees, for the other slow-roll parameter $r$, the tensor-to-scalar ratio, a value of
\begin{eqnarray}
r=\frac{12}{N^2}
\end{eqnarray}
as in  Starobinsky inflation \cite{star} and Higgs Inflation \cite{higgs}, seems to be favored  by the Planck collaboration, which reports 
\begin{eqnarray}
r<0.08,
\end{eqnarray}
whereas chaotic models \cite{chaotic} like the quadratic one, predicts typically
\begin{eqnarray}
 r=\frac{8}{N}\approx 0.2, 
 \end{eqnarray} 
 and are favored by BICEP2. 
The slow-roll parameters $\epsilon,\eta$ and the number of e-folds $N(\sim 50-60)$ during inflation \cite{lr} are defined in terms of the scalar potential $V(\phi)$ of a canonically normalized inflaton field with Lagrangian
\begin{eqnarray}
{\cal L}={\cal L}_E-\frac{1}{2}g^{\mu\nu}\del _\mu \phi \del _\nu \phi-V(\phi)
\end{eqnarray}
as 
\begin{eqnarray}
\epsilon=\frac{1}{2}\left(\frac{V'}{V}\right)^2\, , ~~~
\eta=\frac{V''}{V}\, , ~~~N=\frac{1}{M_P^2}\int_{\phi_{end}}^{\phi_{init}}\frac{V}{V'}d\phi\, .
\label{sr}
\end{eqnarray}

In the next section, we will recall some well known matter from Cosmology, and, in particular, the description of the inflaton as a component of a cosmological perfect fluid. In section 3. we  describe the Starobinsky model. In section 4. we  present the supergravity embedding of the Starobinsky and chaotic inflation in the two different formulations of the ${\cal N}=1$ supergravity. In addition, we  show how integrating out the sgoldstino multiplet, the Volkov-Akulov-Starobinsky supergravity emerges. Finally, section 5 contains our conclusions.  
\section{A (short) Introduction to Cosmology}

The Standard Model for Cosmology describes the Universe as made of different forms of energy densities acting as sources of the gravitational field. The latter in turn is described by a Friedmann-Lema\^{i}tre-Robertson-Walker geometry (FLRW) with metric 
\begin{eqnarray}
ds^2=-dt^2+a(t)^2\left(\frac{dr^2}{1-k r^2}+r^2 d\Omega^2\right)\, ,
\end{eqnarray}
where the three-dimensional constant-time slice is a maximally symmetric space, the Riemann tensor of which satisfies (for constant t)
\begin{eqnarray}
 R_{ijkl}=\frac{k}{a^2}(g_{ik}g_{jl}-g_{il}g_{jk})\, , ~~~k=0,\pm 1\, .
 \end{eqnarray} 
For $k>0,k<0$ and $k=0$ we refer to closed, open or flat Universe, respectively. $a(t)$ is the ``scale factor" which tells us how big is the 3D slice at (comoving) time $t$. The above assumptions are motivated by the ``Copernican Principle", namely that our Universe looks isotropic and homogeneous. Isotropy says that space looks the same in any direction and homogeneity that the metric looks the same everywhere. If a space is isotropic everywhere, then it is homogeneous as well. If we have isotropy 
and homogeneity, then the 3D slice is a maximally symmetric space, i.e., 
\begin{eqnarray}
a)~~ \frac{SO(4)}{SO(3)}~~(k>0),~~~b) ~~\frac{SO(3,1)}{SO(3)}~~(k<0),~~~
c) ~~E_3~~(k=0)\, .
\end{eqnarray}

The Einstein equations are written as 
\begin{eqnarray}
G_{\mu\nu}=R_{\mu\nu}-\frac{1}{2}g_{\mu\nu} R=\kappa^2 T_{\mu\nu} \, ,
~~~(R={R_\mu}^\mu
)\, ,
\end{eqnarray}
where $T_{\mu\nu}$ is the energy-momentum (stress energy) tensor, or, equivalently,
\begin{eqnarray}
R_{\mu\nu}=\kappa^2\left(T_{\mu\nu}-\frac{1}{2}g_{\mu\nu}T\right) \, ,
\end{eqnarray}
which implies
\begin{eqnarray}
T={T_\mu}^\mu=-\frac{1}{\kappa^2} R\, .
\end{eqnarray}
Note that if the energy-momentum tensor is that of a vacuum energy, 
\begin{eqnarray}
T_{\mu\nu}=-\Lambda g_{\mu\nu} \label{ve}
\end{eqnarray}
where $\Lambda=const.,$ by the energy-momentum conservation 
\begin{equation}
\nabla_\mu T^{\mu\nu}=0, \label{ce}
\end{equation}
 we get that
\begin{eqnarray}
R=4\kappa^2 \Lambda=const. 
\end{eqnarray}
and we have a maximally symmetric space-time, which is de Sitter (DS) ($\Lambda>0$),
anti-de Sitter (AdS) ($\Lambda<0$) or Minkowski ($\Lambda=0$). But what is the general form of the  energy-momentum tensor $T_{\mu\nu}$? We can answer to this question if we make the picture of the Universe as being made by perfect fluids, described by an energy density $\rho$ and a pressure $p$. In this case, the energy-momentum tensor can be expressed as 
\begin{eqnarray}
T_{\mu\nu}=(\rho+p)U_\mu U_\nu+p g_{\mu\nu}=(\rho+p)\left(U_\mu U_\nu+
\frac{1}{4}g_{\mu\nu}\right)+\frac{1}{4}(3p-\rho)g_{\mu\nu}\, ,
\end{eqnarray}
where $U^\mu$ is the relativistic four-velocity vector. We may impose an equation of state 
\begin{eqnarray}
p=w\, \rho\, ,
\end{eqnarray}
where $p=p(t)$, $\rho=\rho(t)$ and $w=const.$, with $|w|\leq 1$ by the dominant energy condition. Moreover, the stress tensor conservation (\ref{ce})
implies
\begin{eqnarray}
\frac{\dot{\rho}}{\rho}=-3(1+w)\frac{\dot{a}}{a}\, ,
\end{eqnarray}
which gives that 
\begin{eqnarray}
\rho\propto a^{-3(1+w)}\, .
\end{eqnarray}
Note in particular that a vacuum energy $T_{\mu\nu}\propto g_{\mu\nu}$ as in Eq. (\ref{ve}) leads to 
\begin{eqnarray}
\rho+p=0, ~~~(w=-1)\, ,
\end{eqnarray}
while a traceless stress tensor ${T_\mu}^\mu=0$ implies 
\begin{eqnarray}
3p-\rho=0, ~~~(w=\frac{1}{3}, ~~{\rm radiation})\, .
\end{eqnarray}
The value of the scalar curvature is 
\begin{eqnarray}
R=-\kappa^2 {T_\mu}^\mu=\kappa^2 (1-3w)\rho
\end{eqnarray}
for a single component perfect fluid,  or
\begin{eqnarray}
R=\kappa^2\sum_i(1-3w_i)\rho_i
\end{eqnarray}
in the case of a multi-component perfect fluid. 
For $w=-1$, we have de Sitter or anti-de Sitter space-time depending on the value of $\rho$ (DS for $\rho>0$ and AdS for $\rho<0$) whereas, $R\geq 0$ for $w\leq \frac{1}{3}$ and $\rho_i\geq 0$. 

Because of the symmetries of the FLRW geometry, the Einstein equations now read 
\begin{eqnarray}
&&G_{00}=\kappa^2 T_{00}\, ,\\
&&G_{ij}=\kappa^2 T_{ij}\, ,
\end{eqnarray}
which  give the equations for the scale factor
\begin{eqnarray}
 &&\left(\frac{\dot{a}}{a}\right)^2=\frac{8\pi G}{3}\rho-\frac{k}{a^2}\, ,\\
 &&\frac{\ddot{a}}{a}=-\frac{4\pi G}{3}(\rho+3p) \, , ~~~(k=\pm 1,~0)\, .
 \end{eqnarray} 
 A dot and a double dot denote first and second derivatives with respect to comoving time $t$. By introducing the Hubble parameter 
 \begin{eqnarray}
 H=\frac{\dot{a}}{a}
 \end{eqnarray}
and noticing that 
\begin{eqnarray}
•\frac{\ddot{a}}{a}=\dot{H}+H^2\, ,
\end{eqnarray}
we can rewrite the Einstein equations as (Friedmann equations) 

\begin{eqnarray}
&& H^2=\frac{8\pi G}{3}\rho-\frac{k}{a^2}\, , \label{Fried}\\
&&\dot{H}=-4\pi G(\rho+p)+\frac{k}{a^2}\, .\label{acc}
\end{eqnarray}
By defining 
\begin{eqnarray}
\rho_k=-\frac{3k}{8\pi G}\frac{1}{a^2}, ~~~w_k=-\frac{1}{3}\, ,
\end{eqnarray}
Friedmann equations (\ref{Fried},\ref{acc}) are written as

\begin{eqnarray}
•&& H^2=\frac{8\pi G}{3}(\rho+\rho_k)\, , \label{Fried1}\\
&&\dot{H}=-4\pi G(1+w)\rho-\frac{8\pi G}{3}\rho_k\, ,\label{acc1}
\end{eqnarray}
or, in the case of a multicomponent perfect fluid

\begin{eqnarray}
•&& H^2=\frac{8\pi G}{3}\sum_i \rho_i\, , \label{Fried2}\\
&&\dot{H}=-4\pi G\sum_i(1+w_i)\rho_i\, ,\label{acc2}
\end{eqnarray}
(where $\rho_i$ includes $\rho_k$). By dividing Eq. (\ref{Fried}) by $H^2$ and defining the density parameter 
\begin{eqnarray}
\Omega=\frac{8\pi G}{3 H^2}\rho\, ,
\end{eqnarray}
we get 
\begin{eqnarray}
\Omega-1=\frac{k}{H^2a^2}\, .
\end{eqnarray}
In addition, if we also define the critical energy density 
\begin{eqnarray}
\rho_{crit}=\frac{8\pi G}{3 H^2}\, ,
\end{eqnarray}
we have that 
\begin{eqnarray}
\Omega=\frac{\rho}{\rho_{crit}}\, .
\end{eqnarray}
Similarly, Eq. (\ref{Fried2}) can be compactly written as

\begin{eqnarray}
\sum_i \Omega_i=1\, ,
\end{eqnarray}
where
\begin{eqnarray}
\Omega_i=\frac{\rho_i}{\rho_{crit}}\, ,
\end{eqnarray}
including $\rho_k$.
The second Friedmann equation (\ref{acc}) can be written in terms of the ``deceleration parameter" 
\begin{eqnarray}
q=-a\frac{\ddot{a}}{\dot{a}^2}=-\frac{1}{H^2}(\dot{H}+H^2)\, ,
\end{eqnarray}
as
\begin{eqnarray}
q=\frac{4\pi G}{3H^2}(\rho+3p)=\frac{4\pi G}{3H^2}(1+3w)\rho\, ,
\end{eqnarray}
or  for several components of energy densities 
\begin{eqnarray}
q=\frac{4\pi G}{3H^2}\sum_i(1+3w_i)\rho_i=\frac{1}{2}\sum_i(1+3w_i)
\Omega_i\, .
\end{eqnarray}
Note that $\Omega_k$ does not contribute to $q$ (since $w_k=-1/3$).  As a result, Einstein equations for an FLRW geometry may be written as

\begin{eqnarray}
\sum_i\Omega_i=1, ~~~q=\frac{1}{2}\sum_i(1+3w_i)\Omega_i\, .
\end{eqnarray}

Values for $w$ include
\begin{eqnarray}
&&w=0 ~~~~~~(\text{dust-baryonic matter})\, ,\\
&&w=-\frac{1}{3}~~~(\text{curvature})\, ,\\
&&w=1~~~~~~(\text{fast roll scalar field})\, .
\end{eqnarray}

Whenever a single term dominates, we say that the Universe is dominated by that component of ($p_i,\rho_i$) or ($w_i=p_i/\rho_i,~\rho_i$). For example,  for $w=0$ we have a matter dominated Universe, for $w=\frac{1}{3}$ a radiation dominated, for $w=-1$ a vacuum-energy dominated and for $w=-\frac{1}{3}$ a curvature dominated Universe. When only one component  dominates, the Friedmann equation

\begin{eqnarray}
\left(\frac{\dot{a}}{a}\right)^2\propto \frac{8\pi G}{3} a^{-3(1+w)}
\end{eqnarray}
 can easily be integrated to give
 
\be a\sim
 \begin{cases}
t^{\frac{2}{3(1+w)}},&w\neq 1\\
e^{H t},&w=-1, ~~~(H=const.)\, .
 \end{cases}
 \ee

A scalar field (inflaton) can be viewed as a perfect fluid component of the Universe with Lagrangian 
\begin{eqnarray}
{\cal L}_{\phi}=-\frac{1}{2}g^{\mu\nu}\del_\mu \phi \del_\nu \phi-V(\phi)\, ,
\end{eqnarray}
and stress energy tensor 
\begin{eqnarray}
T_{\mu\nu}^{\phi}=-\frac{\del {\cal L}_\phi}{\del \del_\mu \phi}
\del_\nu \phi+g_{\mu\nu}{\cal L}_\phi\, .
\end{eqnarray}
In components, we find for the latter
\begin{eqnarray}
&&T_{00}^\phi=\frac{1}{2}\dot{\phi}^2+V(\phi),\\
&&T_{ij}^\phi=\left(\frac{1}{2}\dot{\phi}^2-V(\phi)\right)g_{ij}\, ,
\end{eqnarray}
which gives, after comparing with the standard form of the stress tensor of a perfect fluid 
\begin{eqnarray}
T_{00}=\rho,~~~T_{ij}=pg_{ij}\, ,
\end{eqnarray} that the energy density and the pressure of the inflaton are

\begin{eqnarray}
\rho=\frac{1}{2}\dot{\phi}^2+V(\phi), ~~~p=\frac{1}{2}\dot{\phi}^2-V(\phi)\, .
\end{eqnarray}
Therefore, we get that 
\begin{eqnarray}
&\rho+p=\dot{\phi}^2, & \rho+3p=2(\dot{\phi}^2-V)\, ,\\
&\rho-p=2V,& {T_\mu}^\mu=\dot{\phi}^2-4V\, ,
\end{eqnarray}
so that we have 
\begin{eqnarray}
&&p=-\rho, ~~~(w=-1)~~\text{for}~~~\dot{\phi}^2\ll V(\phi)\, ,\\
&&p=\rho, ~~~~~(w=1)~~~~~\text{for}~~~\dot{\phi}^2\gg V(\phi)\, ,
\end{eqnarray}
wheras, in general

\begin{eqnarray}
\frac{1}{2}\dot{\phi}^2(1-w)=(1+w)V\, .
\end{eqnarray}

The matter equations

\begin{eqnarray}
•\frac{\delta {\cal L}_\phi}{\delta \phi}-\del_\mu \frac{\delta {\cal L}_\phi}{\delta \del_\mu \phi}=0\, , 
\end{eqnarray}
together with the Einstein equations are explicitly written as

\begin{eqnarray}
&&H^2+\frac{k}{a^2}=\frac{8\pi G}{3} \left(\frac{1}{2}\dot{\phi}^2+V(\phi)\right)\, ,\label{w1}\\
&&\dot{H}-\frac{k}{a^2}=-4\pi G \dot{\phi}^2\, ,\label{w2}\\
&& \ddot{\phi}+3H\dot{\phi}+V_\phi=0\, . \label{w3}
\end{eqnarray}

Inflation claims to solve the flatness and horizon problems. The flatness problem is the explanation for having today $\Omega\sim 1$ 
without fine-tuning of $|\Omega-1|$ close to zero at early times,
whereas, the horizon problem is the problem of homogeneity of the observed Universe arising form seemingly non-causally connected initial regions. Inflation occurs as long as 
\begin{eqnarray}
\dot{\phi}^2\ll V(\phi),  ~~~|\ddot{\phi}|\ll |3H\dot{\phi}|,|V_\phi|\, ,
\end{eqnarray}
where the potential energy dominates the kinetic energy for sufficient period. The inflationary regime is usually parametrized by the slow-roll parameters
$(\epsilon,\eta)\ll 1$ defined in Eq. (\ref{sr}).

Let us also note at this point that the number of e-foldings $N$ is defined as 
\begin{eqnarray}
N=\int_{t_i}^{t_f} H(t) dt=\ln \frac{a_f}{a_i}, \label{nn}
\end{eqnarray}
where $t_i(t_f)$ are some initial(final) time and $a_{i,f}=a(t_{i,f})$. The definition  (\ref{nn}) reduces to that of eq.(\ref{sr}) once the field equations  (\ref{w1}-\ref{w3}) are used in the slow-roll approximation. Therefore we have that 
\begin{eqnarray}
  a_f=a_ie^{N}=a_i \prod_m e^{N_m}\, ,~~~  \text{for}~~~N=\sum_m N_m\, .
  \end{eqnarray}  
In addition, for a de Sitter background with  
\begin{eqnarray}
H(t)=H_0=const.
\end{eqnarray}
we have
\begin{eqnarray}
H_0 \Delta t =N\, .
\end{eqnarray}

\section{The Starobinksy Model}

The Starobinsky model is the $R+R^2$ theory. It is dual (conformally equivalent) to standard gravity  coupled to a scalar field \cite{Whitt}
with a potential giving rise to inflation. Indeed, the Starobinsky Lagrangian is (in $M_P^{-2}=8\pi G =1$ units)

\begin{eqnarray}
{\cal L}=-\frac{1}{2}R+  \ta  R^2  \, .\label{s}
\end{eqnarray}
By introducing new fields $\sigma$ and $\Lambda$, the Lagrangian (\ref{s}) can be written equivalently as

\begin{eqnarray}
 {\cal L}=-\frac{1}{2}R+\frac{1}{2}\sigma(\Lambda-R)+\ta \Lambda^2\, , \label{s1}
 \end{eqnarray} 
where the field $\sigma$ is a Langrange multiplier which enforces the constraint $\Lambda=R$. We may write (\ref{s1}) as
\begin{eqnarray}
 {\cal L}=-\frac{1}{2}R(1+\sigma)+\frac{1}{2}\sigma\Lambda+\ta \Lambda^2\, , \label{s10}
 \end{eqnarray} 
where we observe that we have a Jordan frame function $(1+\sigma)$. Going to the Einstein frame through the change of variables 
\begin{eqnarray}
g_{\mu\nu}'=g_{\mu\nu}(1+\sigma)^{-1}\, ,
\end{eqnarray}
we get 
that the Lagrangian (\ref{s10}) is written as

\begin{eqnarray}
 {\cal L}=-\frac{1}{2}R-\frac{1}{2}\del_\mu\phi\del^\mu \phi-\frac{1}{16 \ta}(1-e^{-\sqrt{\frac{2}{3}}\phi})^2\, ,\label{s11}
 \end{eqnarray} 
where the field $\phi$, the ``scalaron" 
is  defined as 
\begin{eqnarray}
1+\sigma=e^{\sqrt{\frac{2}{3}}\phi}\, .
\end{eqnarray}
In Fig.1 below, the scalaron potential 
\begin{eqnarray}
V=\frac{1}{16 \ta}(1-e^{-\sqrt{\frac{2}{3}}\phi})^2\, ,
\end{eqnarray}
has been plotted. At $V_\phi=\del V/\del \phi=0$, supersymmetry is unbroken whereas during the inflationary phase (``de Sitter plateau"), we have $\del V/\del \phi\neq 0$ and supersymmetry is broken. 
\begin{figure}[!htb]
\centering
\includegraphics[scale=.7]{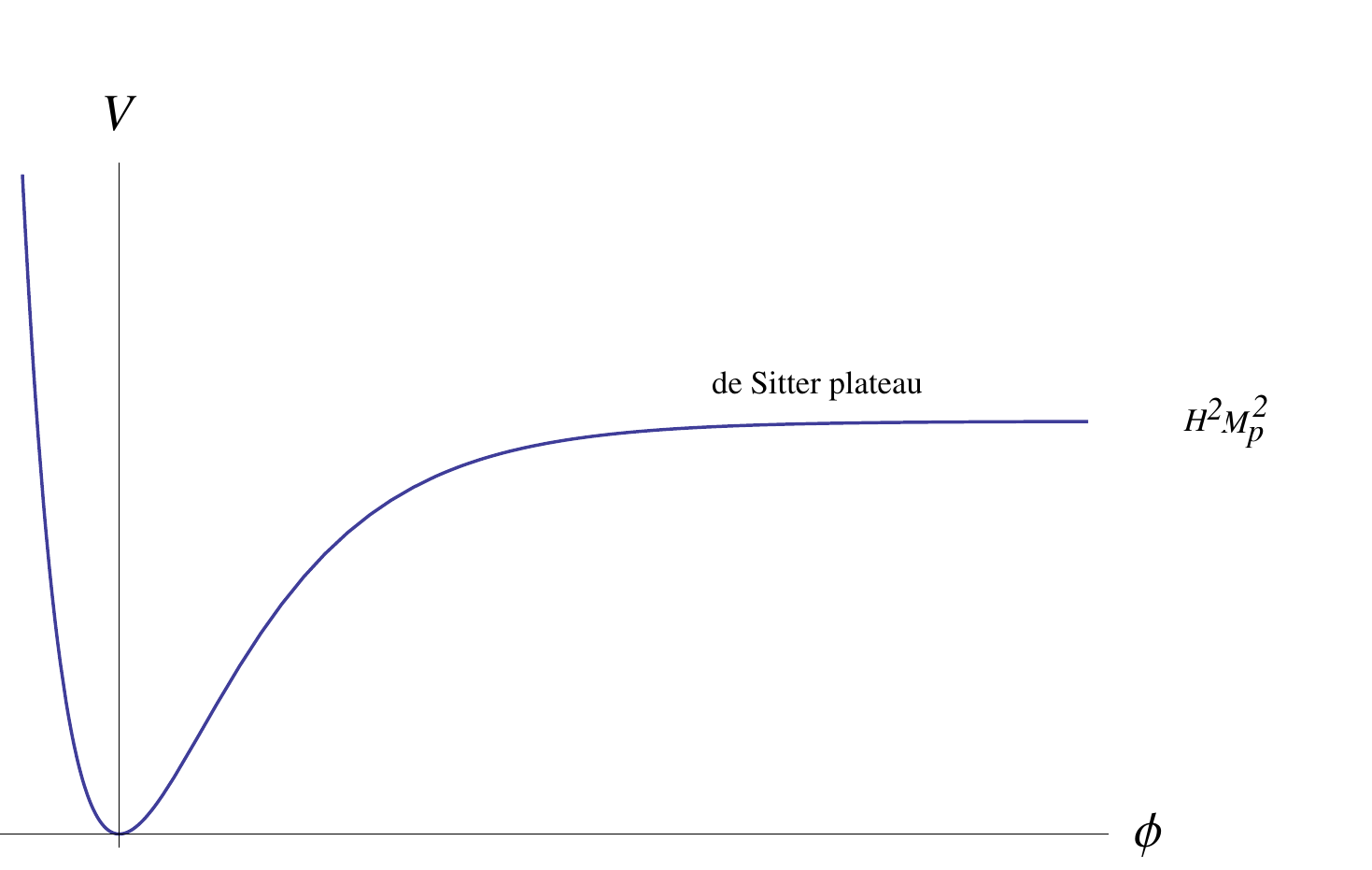}
\caption{The ``scalaron" potential in $R+R^2$ theory.}
\label{fig:digraph}
\end{figure}

In trying to describe the Starobinsky model in supergravity, let us note that the off-shell components of the gravity field give rise to extra massive modes in higher curvature supergravity. For example let us consider a Lagrangian quadratic to curvature of the form

\begin{align*}
{\cal L}&={\cal L}_E+{\cal L}_{R^2}+{\cal L}_{Weyl^2}\nonumber \\
&=\frac{1}{2\kappa^2} R+\ta R^2+\text{\textbeta} \,W_{\mu\nu\rho\sigma}^2\, .
\end{align*}
The graviton $g_{\mu\nu}$ has a total of six degrees of freedom, since the total number of the ten independent components of $g_{\mu\nu}$ is reduced by the number four, which  the number of diffeomorphisms

\begin{eqnarray}
10~-\underbrace{4}_{
          \mathclap{\text{ diffeomorphisms}}}=6\, .
\end{eqnarray}
These six degrees of freedom of $g_{\mu\nu}$ give rise to a scalar degree of freedom with mass $m_0$ and a massive spin-2 state with mass $m_2$, according to the splitting 

\begin{eqnarray}
•6=\underbrace{1}_\text{scalar}+\underbrace{5}_\text{spin 2}
\end{eqnarray}
where 
\begin{eqnarray}
m_0^2\sim \frac{1}{\kappa^2 \ta}\, , 
~~~m_2^2\sim -\frac{1}{\kappa^2 \text{\textbeta}}\, .
\end{eqnarray}
Therefore, the massive spin-2 state is a tachyon ($\text{\textbeta}>0$) or a ghost ($\text{\textbeta}<0$). This state decouples in the $\beta\to 0$ limit, leading to  Starobinsky $R+R^2$ theory with dynamical degrees of freedom a massless graviton and a scalar field. 

The higher curvature theory should be compared with the chaotic model \cite{chaotic}. The Lagrangian of the latter is 
\begin{eqnarray}
{\cal L}=-\frac{1}{2}R-\frac{1}{2}(\partial_\mu \phi)^2-\frac{1}{2}m^2
\phi^2\, ,
\end{eqnarray}
where the mass of the inflaton is $m\sim H$. It can be seen that during inflation, the de Sitter cosmological constant $\Lambda$  is 
\begin{eqnarray}
\Lambda\sim H^2M_P^2\, ,
\end{eqnarray}
where $H/M_p\sim 10^{-5}$ for both the higher curvature and the chaotic model.

\section{Supergravity Embedding}

Here we present the supergravity embedding of these two models which is minimal in two respects: It uses the minimal set of multiplets needed to describe these models. It also uses the minimal off-shell representations of the underlying local supersymmetry algebra. The latter introduces new fields which are ``auxiliary"
(not propagating) in the standard Einstein supergravity but become propagating when higher curvature terms are introduced. The minimal supergravity extension of such a model was derived in the late eighties \cite{Cecotti1,CFPS} in two different forms depending of two different off-shell completion of the supergravity multiplet
\begin{eqnarray}
  &&a) ~~~V_\mu^a,~~\psi_\mu,~~A_\mu,~~S,~~P\, ,\\
  &&b)~~~V_\mu^a,~~\psi_\mu,~~A_\mu,~~b_{\mu\nu}, ~~~~~A_\mu\to A_\mu+\del_\mu a,~~b_{\mu\nu}\to b_{\mu\nu}+\del_\mu \xi_\nu-\del_\nu \xi_\mu\, .
  \end{eqnarray}  
  The first case $(a)$ is the old-minimal supergravity and the second case $(b)$ is the new-minimal one. 
  The six bosonic degrees of freedom which make the gravity multiplet to have the same number of bosons and fermions ($12b+12f)$ give two different supergravity extensions of the Starobinsky model.
  
  The off-shell components of (${\cal N}=1$) supergravity fields give extra massive supermultiplets in higher curvature supergravity. 
In the old-minimal supergravity we have 
\begin{eqnarray}
&&\text{graviton}~~~g_{\mu\nu}: ~~~6=1_0+5_2, ~~~
A_\mu,~~S,~~P: ~~~6=3_1+1_0+1_0+1_0\, ,\\
&&\text{gravitino}~~~\psi_\mu: ~~~16-4=12=2\times \frac{3}{2}+2\times\frac{1}{2}\, ,
\end{eqnarray}
which describes two massive chiral multiplet  $2(\frac{1}{2},2(0))$ with ($4b+4f)$ and a ghost spin-2 multiplet $(2,2(\frac{3}{2}),1)$. In the new-minimal supergravity, the $3+3$ degrees of freedom of the gauge fields 
$A_\mu,b_{\mu\nu}$ fill the bosonic part of a physical massive vector multiplet $(1,2(\frac{1}{2}),0)$ \cite{fv}. 

The ``dual" standard supergravity action contains, in the $a)$ formulation two ``matter" chiral (massive) multiplets $T,S$ ($4b+4f$) while in the $b)$ formulation contains a ``massive" vector (or tensor) multiplet ${\cal V}$ ($4b+4f$). The main difference is that in the $a) $ theory we are in a presence of a ``four-field"  model   whereas in the $b)$ theory we have a ``single-field" inflaton model since the other three bosonic degrees of freedom combine in a massive vector.

Standard supergravity formulae allow to describe the $a)$ theory in terms of a K\"ahler potential $K$ and a superpotential $W$. It turns out that their form is 

\begin{eqnarray}
K=-3 \log\Big{(}1+T+\bar T-h(S,\bar S)\big{)}\, , ~~~ 
W=\lambda \, T\,  S\, ,\label{kw}
\end{eqnarray}
where $\lambda$ is a constant related to the $\ta$ parameter and 
$h(S,\bar S)$ is an arbitrary real function which starts with 

\begin{eqnarray}
h(S,\bar S)=S\bar S+{\cal O}(S^3)\, .
\end{eqnarray}
It is possible to choose the function $h(S,\bar S)$ appropriately in order  to make the inflationary trajectory stable. Here the ``inflaton" is identified with the $Re T$, the real part of the scalar $T$, while the other three scalars are ``extremized". The potential for 
\begin{eqnarray}
Re T=e^{-\sqrt{\frac{2}{3}}\phi}
\end{eqnarray}
is the Starobinsky potential. It can be shown that this theory, for any $h(S,\bar S)$ is ``dual" to a higher curvature supergravity theory. The scalar supercurvature ${\cal R}$ is a chiral superfield \cite{FZ}
\begin{eqnarray}
\bar D _{\dot{a}} {\cal R}=0\, ,
\end{eqnarray}
and $h(S,\bar S)$  corresponds to terms of the form $h({\cal R},\bar {\cal R})$ in the supergravity side. It is important to notice that the inflaton potential is an ``F-term" potential, which means that it comes from the standard expression 

\begin{eqnarray}
V(T,S)=e^{K}\left(D_iW \, D_{\bar{j}}\bar W K^{i\bar j}-3 |W|^2\right) , ~~~(i,j=S,T)\, .
\end{eqnarray}
The inflaton potential is then

\begin{eqnarray}
V(\phi)=V(T,S)\Big{|}_{\frac{\del V}{\del S}=0, 
~\frac{\del V}{\del ImT}=0} \, .
\end{eqnarray}

It happens that all supersymmetric models for the inflaton potential considered in the literature 
\cite{Kallosh-Linde,ENO,FKLP,FKR,FKvP,KLW,KLR,nano,FKRim,KLr,jap,E11,E2} are mostly deformations of the previous model with modification of $K(T,\bar T,S,\bar S)$ and of $W(T,S)$ but still keeping the same ($S,T$) chiral multiplet content.

It is possible to show that at least two multiplets are needed to get an inflationary potential. In fact former theories with higher supercurvature terms of F-term type with chiral function $f({\cal R})$ ($\bar D f=0$) were considered in the past \cite{KT} but were shown \cite{ENO,FKP} not to produce an inflationary potential. 

An important deformation of the $(S,T)$ model from which the concept of ``attractor" \cite{KLR} came from,  is a superpotential of the type

\begin{eqnarray}
  W(S,T)=S f(T) \, ,\label{w}
  \end{eqnarray}  
which allows bosonic potentials containing arbitrary functions of the inflaton $f(\tanh \frac{\phi}{\sqrt{6}})$. These theories are no longer equivalent to pure higher curvature supergravity but in certain cases, to higher curvature coupled to a (single) chiral multiplet. For instance, taking $K$ as in (\ref{kw}) but $W$ now as (\ref{w}), the dual higher derivative supergravity is a matter coupled theory with \cite{CK}
\begin{eqnarray}
\Phi=e^{-\frac{1}{3}K}=1+T-\frac{f(T)}{f'(T)}+\bar T -\frac{\bar{f}(\bar{T})}{\bar{f}'(\bar{T})}
\end{eqnarray}
and  a term 
\begin{eqnarray}
\frac{1}{|f'(T)|^2}{\cal R }\bar {\cal R}\, .
\end{eqnarray}
Both terms become $T$-independed if $f(T)=a\,  T$.

The $b)$ formulation gives directly a single-field inflation model where a ``D-term" potential for the massive superfield is generated. The most general self-interaction of such massive vector multiplet with spin content 
$(1,2(\frac{1}{2}),0)$ resides on a real function $J$ of a real variable $C$: $J(C)$ \cite{vP}
The bosonic part of the supergravity action is 

\begin{eqnarray}
{\cal L}=-\frac{1}{2} R-\frac{1}{4}F_{\mu\nu}(B) F^{\mu\nu}(B)+\frac{g^2}{2}
J''(C)B_\mu B^\mu+\frac{1}{2}J''(C) \del_\mu C\del^\mu C-\frac{g^2}{2}
J'(C)^2 \, ,\label{J}
\end{eqnarray}
so that the potential is 
\begin{eqnarray}
V(C)=\frac{g^2}{2}J'(C)^2\, .
\end{eqnarray}
The equation (\ref{J}) actually coincides with the self-interaction of a massive vector multiplet in global supersymmetry \cite{fayet}.
Note also that (\ref{J}) depends only on $J',J''$, so a linear term in $J$ shifts $J'$ by  a constant but leaves $J''$ invariant. This constant is the so-called Fayet-Iliopoulos (FI) term. By using the St\"uckelberg trick one writes (\ref{J}) as a gauge theory by shifting 
\begin{eqnarray}
A_\mu=B_\mu-\frac{1}{g}\del_\mu a\, ,
\end{eqnarray}
so that 
\begin{eqnarray}
\frac{g^2}{2}J''(C)B_\mu B^\mu=\frac{g^2}{2}J''(C)\Big{(}A_\mu+\frac{1}{g}\del_\mu a\Big{)}^2\, .
\end{eqnarray}
In the limit $g\to 0$ the theory becomes
\begin{eqnarray}
{\cal L}=-\frac{1}{2} R-\frac{1}{4}F_{\mu\nu}(A) F^{\mu\nu}(A)+\frac{1}{2}
J''(C)\Big{(}\del_\mu a \del^\mu a+\del_\mu C\del^\mu C\Big{)}^2 
\, .\label{J2}
\end{eqnarray}
The $(a,C)$ variables can be complexified to $T=- C+i\sqrt{\frac{2}{3}}\, a$ and the $J$-function can be interpreted as a K\"ahler potential 
\begin{eqnarray}
J=-\frac{1}{2}K(Re T)\, .
\end{eqnarray}

The higher curvature supergravity in the $b)$ formulation is ``dual" to a self-interacting massive vector multiplet with a very precise choice \cite{FKLP} of 
\begin{eqnarray}
J(C)=\frac{3}{2}\Big{(}\log (-C)+C\Big{)}\, .
\end{eqnarray}
Computation of the potential, for a canonically normalized field 
\begin{eqnarray}
C=-e^{\sqrt{\frac{2}{3}}\phi}\, ,
\end{eqnarray}
leads to Starobinsky potential and Lagrangian \cite{FKR,FKLP}
\begin{eqnarray}
{\cal L}=\cdots-\frac{1}{2}(\del_\mu\phi)^2-\frac{9}{8}g^2(1-e^{-\sqrt{\frac{2}{3}}\phi})^2\, ,
\end{eqnarray}
so that the supersymmetric generalization just reproduces the single-field Starobinsky model with $\ta\propto g^{-2}$. It is interesting to observe that the particular form of $J(C)$ corresponds to an $SU(1,1)/U(1)$ symmetric K\"ahler manifold with a parabolic isometry being gauged. For a 
K\"ahler potential 
\begin{eqnarray}
K=-3 \alpha \log Re \,T \, ,
\end{eqnarray}
the curvature is 
\begin{eqnarray}
R(C)=\frac{J'''(C)^2-J''(C) J^{IV}(C)}{2J''(C)^2}=-\frac{2}{3\alpha}
\end{eqnarray}
and for $\alpha\to \infty$, the curvature vanishes $R(C)\to 0$. The $\alpha$-depended potential becomes \cite{FKLP}
\begin{eqnarray}
V(\phi)=\frac{9}{8}g^2(1-e^{-\sqrt{\frac{2}{3\alpha}}\phi})^2=\frac{9}{8}g^2 P(C)^2\, ,
\end{eqnarray}
where $P(C)=J'(C)$. 
Note that the canonical variable $\phi$ is related to the $C$ variable by the equation  
\begin{eqnarray}
J''(C)=-\left(\frac{d\phi}{d C}\right)^2=P'(C)\, .
\end{eqnarray}
It follows then, with $P(\phi)=P(C(\phi))$, $P'(C)=dP/dC$ and $P'(\phi)=dP/d\phi$
\begin{eqnarray}
P'(C)=P'(\phi)\frac{d\phi}{dC}=-\left(\frac{d\phi}{d C}\right)^2\, ,
\end{eqnarray}
so that 
\begin{eqnarray}
P'(\phi)=-\frac{d\phi}{d C}
\end{eqnarray}
and 
\begin{eqnarray}
&&C(\phi)=\int d\phi \frac{dC}{d\phi}=-\int d\phi \frac{1}{P'(\phi)}\, ,\\
&&J(C)=\int dC J'(C)=\int P(\phi)\frac{dC}{d\phi} d\phi =-\int \frac{P(\phi)}{P'(\phi)}d\phi\, .
\end{eqnarray}
In addition, the curvature in the $\phi$-variable is
\begin{eqnarray}
R(\phi)=-4\frac{P'''(\phi)}{P'(\phi)}\, ,
\end{eqnarray}
whereas, 
the kinetic term of the K\"ahler manifold is \cite{FFS}
\begin{eqnarray}
\frac{1}{2}
J''(C)\Big{(}\del_\mu a \del^\mu a+\del_\mu C\del^\mu C\Big{)}^2
=-\frac{1}{2}	\Big{[}(\del _\mu \phi)^2+P'(\phi)^2(\del_\mu a)^2
\Big{]}\, .
\end{eqnarray}
The previous equations allow us to compute $C(\phi)$ once 
$P'(\phi)=-d\phi/dC$ is solved. 

The one-field supergravity model for inflation can be deformed in two ways:
\begin{enumerate}
\item Simply change $J(C)$, i.e., change the K\"ahler manifold. 
\item Do not change the manifold but change its gauged isometry.
\end{enumerate}
For the case of symmetric spaces, this procedure generates five models. Three with constant curvature depending whether  a parabolic, elliptic or hyperbolic isometry is gauged and two with vanishing curvature where the parabolic or elliptic isometry is gauged \cite{FFS}. 

\subsection{Chaotic Inflation}

With an ``F-term" multi-field potential term, it is hard to obtain (at most in some directions of the field space) a quadratic potential. One way is to impose a shift symmetry on the K\"ahler potential \cite{KYY,KLW,E11,E2}.
In terms of the $(T,S)$ chiral fields this exchange the role of $(ImT,ReT)$ since it is now the $Im T$ which plays the role of the inflaton. It is then natural, in the supergravity dual to cal this scenario ``imaginary Starobinsky model" \cite{FKRim} even if a coupling to matter is needed in order to stabilize the $ReT$ component. 

In the case of chaotic inflation in the $b)$ single-field supergravity formulation, an exact model is possible since we can take a flat K\"ahler space where we gauge a parabolic isometry (translations). The alternative gauging of an elliptic isometry would give a quartic potential. For this case,
\begin{eqnarray}
J''(C)=const. , ~~~J(C)=-\frac{m^2}{2}C^2+\xi C\, ,
\end{eqnarray}
but the FI term is irrelevant. Then 
\begin{eqnarray}
P(\phi)=\phi, ~~~\text{and}~~~V(\phi)=\frac{1}{2}m^2 \phi^2\, .
\end{eqnarray}
This model can also be obtained from the constant curvature case by taking the limit \cite{KLR}
\begin{eqnarray}
\alpha\to \infty, ~~g^2\to \infty, ~~~\text{with}~~~m^2\propto \frac{g^2}{\alpha}~~\text{fixed}\, ,
\end{eqnarray}
so that 
\begin{eqnarray}
g^2(1-e^{-\sqrt{\frac{2}{3\alpha}}\phi})^2\to \frac{1}{2}m^2 \phi^2\, .
\end{eqnarray}

\subsection{Integrating out the sgoldistino multiplet: The Volkov-Akulov-Starobinsky Supergravity}

We observe that the above potential is a $D^2$ term so during inflation $D$ is large and the gaugino is the goldstino. The decoupling of the other (chiral) component occurs when $J''(C)\to 0$ and we get an unbroken gauge symmetry in de Sitter space (Freedman model \cite{Freed}). Supergravity can be formulated in different conformal gauges (different Jordan functions).
In the type $a)$
formulation of inflation \cite{Kallosh-Linde}, there are three basic fields (chiral superfields), the conformon multiplet $S_0$, which is not physical, the scalaron multiplet $T$, which contains the inflaton and the goldstino multiplet $S$ which contains the sgoldstino. The latter is the goldstino partner and it is  just 
$S|_{\theta=0}$ as supersymmetry is linearly realized. The superpotential 
$W=S T$ has F-terms
\begin{eqnarray}
  &&\frac{\del W}{\del S}=T\neq 0~~~\text{during inflation}\, ,\\
  &&\frac{\del W}{\del T}=S=0~~~\text{during inflation and later}\, .
  \end{eqnarray}  
In fact, this explains why in the $a)$ formulation two chiral fields are needed. 

A new efective Lagrangian, since supersymmetry is badly broken during inflation, can be obtained replacing the sgoldstino multiplet $S$ by the Volkov-Akulov superfield $X$ which satisfies the constraint
\begin{eqnarray}
X^2=0
\end{eqnarray}
and allows to express $X$ in terms of the golstino $G_\alpha$ as \cite{goldstino}
\begin{eqnarray}
X=\frac{G_\alpha G^\alpha}{2 F_X}+\sqrt{2}\theta^\alpha G_\alpha+\theta^\alpha\theta_\alpha F_X\, .
\end{eqnarray}
In the dual supergravity theory, this corresponds to the chiral scalar supercurvature ${\cal R}$ to become nilpotent \cite{ADFS}
\begin{eqnarray}
{\cal R}^2=0\, .
\end{eqnarray}

Let us recall that the Volkov-Akulov Lagrangian is 
\begin{eqnarray}
{\cal L}_{VA}=f^2\det V_{\alpha \mu}\, , \label{va1}
\end{eqnarray}
where $f$ is the SUSY breaking parameter and 
\begin{eqnarray}
V_{\alpha\mu}=\delta_{\alpha\mu}+\frac{i}{f^2} \bar G 
\gamma_\alpha \del _\mu G\, .
\end{eqnarray}
Supersymmetry is non-linearly realised and  eq. (\ref{va1}) is invariant under the transformation
\begin{eqnarray}
\delta G=f\epsilon+\frac{i}{f} \bar G \gamma^\mu \epsilon \del _\mu G\, .
\end{eqnarray}
In superspace, we can write 
\begin{eqnarray}
{\cal L}=X\bar X\Big{|}_D+fX\big{|}_F, ~~~~(X^2=0)\, .
\end{eqnarray}
When coupled to supergravity, one gets a theory of a massive gravitino coupled to gravity with K\"ahler potential and superpotential
\begin{eqnarray}
K=-3\log(1-X\bar X)=3 X\bar X,~~~W=fX+W_0\, .
\end{eqnarray}
The vacuum energy and the gravitino mass turn out to be
\begin{eqnarray}
V_0=\frac{1}{3}|f|^2-3 |W_0|^2, ~~~m_{3/2}=|W_0|\, ,
\end{eqnarray}
whereas the Noether current is 
\begin{eqnarray}
J_{\alpha \mu}\sim f \gamma_\mu G+\cdots
\end{eqnarray}
Now we may couple V-A to supergravity and to the scalaron multiplet. The massive  spin 3/2 action is of the form 
\begin{eqnarray}
\frac{1}{2\kappa^2}R+\psi\del \psi +m_{3/2} \psi\sigma\psi +\kappa^2 \psi^4-V(f,m_{3/2})\, .
\end{eqnarray}
The constrained V-A superfield $X$ is then coupled to the scalaron leading to an almost standard supergravity with k\"ahler potential and superpotential
\begin{eqnarray}
K=-3\log(T+\bar T-X\bar X), ~~~W=MXT+fX+W_0, ~~~X^2=0\, .
\end{eqnarray}
The above data give rise to a potential (with no-scale structure 
$V\geq 0$) for the scalar $T$
\begin{eqnarray}
V=\frac{|MT+f|^2}{3(T+\bar T)^2}\, .
\end{eqnarray}
By defining 
\begin{eqnarray}
T=e^{\sqrt{\frac{2}{3}}\phi}+ia \sqrt{\frac{2}{3}}\, ,
\end{eqnarray}
one gets \cite{ADFS}
\begin{align}
{\cal L}&=\frac{1}{2}R-\frac{1}{2}(\del \phi)^2-\frac{M^2}{12}
(1-e^{-\sqrt{\frac{2}{3}}\phi})^2-\frac{1}{2}e^{-2\sqrt{\frac{2}{3}}\phi}
(\del a)^2-\frac{M^2}{18}e^{-2\sqrt{\frac{2}{3}}\phi}a^2\nonumber \\
&={\cal L}_{Starobinsky}+{\cal L}_{axion}+(\text{fermionic terms})\, .
\label{sa}
\end{align}

The dual supergravity action is 
\begin{eqnarray}
{\cal L}(S_0,{\cal R})=-[S_0\bar S_0-\frac{{\cal R}\bar {\cal R}}{M^2}]_D
+(W_0+\xi \frac{{\cal R}}{S_0})S_0^3+\sigma {\cal R}^2 S_0\, .
\end{eqnarray}
The bosonic part of this action can be obtained by dualizing the previous action (\ref{sa}) having set
\begin{eqnarray}
e^{\sqrt{\frac{2}{3}}\phi}=1+2\chi\, ,
\end{eqnarray}
and Weyl rescaling 
\begin{eqnarray}
g_{\mu\nu}\to (1+2 \chi)g_{\mu\nu}\, .
\end{eqnarray}
The result is the Lagrangian
\begin{eqnarray}
{\cal L}=\frac{1}{2}(1+2\chi)R-\frac{1}{2}\frac{(\del a)^2}{1+2 \chi}
-\frac{M^2}{3}(\chi^2+\frac{a^2}{6})\, ,
\end{eqnarray}
which can be written equivalently as 
\begin{eqnarray}
{\cal L}=\frac{1}{2}(1+2\chi)R-\frac{M^2}{18}a^2+A^\mu \del_\mu a+
\frac{1}{2}(1+2\chi)A_\mu^2 \, .
\end{eqnarray}
The dual Lagrangian is then obtained by integrating over $a$ and $\chi$. The result is ($A_\mu\to \sqrt{\frac{2}{3}}A_\mu$)
\begin{eqnarray}
{\cal L}=\frac{1}{2}\left(R+\frac{2}{3}A_\mu^2\right)+\frac{3}{4M^2}
\left(R+\frac{2}{3}A_\mu^2\right)^2+\frac{3}{M^2}(\nabla^\mu A_\mu)^2\, .
\end{eqnarray}
This is the $R+R^2$ Lagrangian with $S=P=0$. Note that the axion field is much heavier than $\phi$ during inflation where $\phi=\phi_0$ is positive and  large
\begin{eqnarray}
 m_\phi^2=\frac{M^2}{9}e^{-2\sqrt{\frac{2}{3}}\phi_0}\ll 
 m_a^2=\frac{M^2}{9}\, .
 \end{eqnarray}

\section{Conclusions}

In these lectures, we have presented the supergravity embedding of higher curvature models of inflation. The prototype of such models is the Starobinksy $R+R^2$ gravity. This theory does not describe only the GR degrees of freedom, i.e. the helicity-2 massless graviton, but in addition it propagates a scalar degree, the ”scalaron”. It also predicts a tiny value for the tensor-to-scalar ratio $r$ due to an addition 1/N suppression with respect to the scalar tilt $n_S$, which is in  perfect agreement with the Planck data.  However, this prediction is in conflict with the BICEP2 results. This tension between Planck and BICEP2 is expected to be resolved soon. We should stress that the inflationary predictions of the Starobinsky model   is identical to leading order with that of  Higgs inflation \cite{higgs}. 
 As it has been shown in \cite{KDR}, this is due to the fact that Higgs inflation and Starobinsky model are identical during slow-roll, where the kinetic terms of the Higgs and the scalaron are subleading with respect to their potentials. We have presented how the bosonic Starobinsky model can be embedded in ${\cal N} = 1$  minimal supergravity. In fact, since it is a higher curvature theory, it is described both in old-minimal  (formulation a)) as well as in new-minimal  ((b) formulation)  
 ${\cal N} = 1$ supergravity. The supergravity formulation of the chaotic model has also been presented.

Concluding, we would like to stress that the Starobinsky model and its descendants as well as their supergravity avatars cannot be excluded as long as BICEP2 results are not independently confirmed.

\vskip.3in
\noindent
{\bf {Acknowledgment}}

\vskip.1in
\noindent
Parts of these lectures are based on collaborations with A. Antoniadis, E. Dudas, P. Fre, R. Kallosh, A. Linde, M. Porrati, A. Riotto, A. Sagnotti, A. Sorin and A. van Proeyen, whom we would like to thank.
This research was implemented
under the ARISTEIA Action of the Operational Programme Education and Lifelong Learning
and is co-funded by the European Social Fund (ESF) and National Resources. It is partially
supported by European Union’s Seventh Framework Programme (FP7/2007-2013) under REA
grant agreement n. 329083. S.F. is supported by ERC Advanced Investigator Grant n. 226455
Supersymmetry, Quantum Gravity and Gauge Fields (Superfields).

\end{document}